# Comment

# History of the Kohlrausch (stretched exponential) function: Focus on uncited pioneering work in luminescence

**M. Berberan-Santos**[1,*], **E. N. Bodunov**[2], and **B. Valeur**[3]

[1]Centro de Química-Física Molecular, Instituto Superior Técnico, Universidade Técnica de Lisboa, 1049-001 Lisboa, Portugal

[2] Physical Department, Petersburg State Transport University, St. Petersburg, 190031, Russia

[3] CNRS UMR 8531, Laboratoire de Chimie Générale, CNAM, 292 rue Saint-Martin, 75141 Paris cedex 03, and Laboratoire PPSM, ENS-Cachan, 61 avenue du Président Wilson, 94235 Cachan cedex, France

*berberan@ist.utl.pt

Important elements for the history of the Kohlrausch (or stretched exponential) relaxation function were recently presented by Cardona, Chamberlin, and Marx [1].

The Kohlrausch function is given by

$$P(t) = \exp\left[-\left(t/\tau_0\right)^\beta\right], \qquad (1)$$

where $P(t)$ is a linear function of a property of a system that evolves towards equilibrium after the sudden removal of a perturbation, $0<\beta\leq 1$, and $\tau_0$ is a parameter with the dimensions of time.

In studies of the relaxation of complex systems, the Kohlrausch function is frequently used as a purely empirical relaxation function, given that it allows gauging in a simple way deviations from the "canonical" single exponential behaviour by means of parameter $\beta$. There are nevertheless theoretical arguments to justify its relatively common occurrence.

The first use of the stretched exponential function to describe the time evolution of a non-equilibrium quantity is usually credited to Rudolph Kohlrausch (1809-1858), who in 1854 [2] applied it to the discharge of a capacitor (Leyden jar), after concluding that a simple exponential of time was inadequate [3].

Like Cardona, Chamberlin, and Marx [1], the present authors also commented on the frequently careless citation of Kohlrausch's work [4,5].

We would like now to draw attention to a set of pioneering works on the Kohlrausch function, one of which published 101 years ago [6], and that are not mentioned in [1]. All these uncited works pertain to the description of luminescence decays, and are briefly discussed in [4] and [5] in connection with the stretched exponential relaxation function.

The Kohlrausch function was most likely used for the first time in luminescence by Werner in 1907 [6], in order to describe the short-time luminescence decay of an inorganic phosphor. This pioneering work of a Ph. D. student of Philipp von Lenard (1862-1947, 1905 Nobel Prize in Physics) [7] in Kiel has received only 8 SCI citations, however one of these comes from a paper by Marsden [8]. The article is also cited in a monograph (where we found it) [9], and in a few more books [10,11]. In his article, Werner does not cite Kohlrausch, and again uses the Kohlrausch function as an empirical function. Werner's work is thus the second documented use of the Kohlrausch function.

In the field of condensed matter luminescence, the Kohlrausch function has firm grounds on several models of luminescence quenching, namely diffusion-controlled contact quenching [12] (75 citations), where the transient term has $\beta=1/2$, and diffusionless resonance energy transfer by the dipole-dipole mechanism, with $\beta=1/6$, $1/3$ and $1/2$ for one-, two- and three-dimensions, respectively, see [13] (1139 citations) and [14] (23 citations). Other rational values of $\beta$ are obtained for different multipole interactions, e.g. $\beta=3/8$ and $\beta=3/10$ for the dipole-quadrupole and quadrupole-quadrupole mechanisms in three-dimensions [15] (1099 citations).

All these works ([6, 12-15]) antedate (in the Werner case by many decades) the 1970 Williams and Watts' paper on dielectric relaxation, and are a significant part of the history of the stretched exponential relaxation function.